\documentclass[aps,prl,preprint,superscriptaddress,nofootinbib,citeautoscript]{revtex4-2}
\setcitestyle{super}

\usepackage{microtype}
\usepackage{newtxtext,newtxmath}
\usepackage[separate-uncertainty=true]{siunitx}
\usepackage{physics}
\usepackage{enumitem}
\usepackage{booktabs}
\usepackage{graphicx}

\begin{document}

\graphicspath{{./figures}}
\title{Breakdown of self-similarity in light transport}

\author{Ernesto Pini}
\affiliation{Department of Physics and Astronomy, Universit\`{a} di Firenze, Sesto Fiorentino, Italy}
\affiliation{European Laboratory for Non-linear Spectroscopy (LENS), Sesto Fiorentino, Italy}

\author{Giacomo Mazzamuto}
\affiliation{National Research Council -- National Institute of Optics (CNR-INO), Sesto Fiorentino, Italy}
\affiliation{European Laboratory for Non-linear Spectroscopy (LENS), Sesto Fiorentino, Italy}
\affiliation{Department of Physics and Astronomy, Universit\`{a} di Firenze, Sesto Fiorentino, Italy}

\author{Francesco Riboli}
\affiliation{National Research Council -- National Institute of Optics (CNR-INO), Sesto Fiorentino, Italy}
\affiliation{European Laboratory for Non-linear Spectroscopy (LENS), Sesto Fiorentino, Italy}

\author{Diederik S. Wiersma}
\affiliation{Department of Physics and Astronomy, Universit\`{a} di Firenze, Sesto Fiorentino, Italy}
\affiliation{European Laboratory for Non-linear Spectroscopy (LENS), Sesto Fiorentino, Italy}
\affiliation{Istituto Nazionale di Ricerca Metrologica (INRiM), Turin, Italy}

\author{Lorenzo Pattelli}
\email{l.pattelli@inrim.it}
\affiliation{Istituto Nazionale di Ricerca Metrologica (INRiM), Turin, Italy}
\affiliation{European Laboratory for Non-linear Spectroscopy (LENS), Sesto Fiorentino, Italy}

\date{\today}

\maketitle

\begin{bf}
Transport processes underpin a multitude of phenomena, ranging from the propagation of atoms on lattices, to the mobility patterns of microorganisms and earthquakes, to name a few. The dynamics of these processes is very rich and key to understanding the complex nature of the underlying physics, but the way in which we classify them is often too simplistic to fully reflect this complexity.
Here, we report on the experimental observation of a breakdown of self-similar propagation for light inside a scattering medium -- a transport regime exhibiting different scaling rates for each spatial moment of the associated probability distribution. Notably, we show that this phenomenon arises for light waves even in the simple case of isotropic and homogeneous disorder, and can be controlled by tuning the turbidity of the system.
These results support the idea that the traditional dichotomy between normal and anomalous diffusion is reductive and that a richer framework should be constructed based on the concept of self-similarity as this class of transport regimes may be far more common that it is currently believed. In addition, this insight can help understand scenarios where transport is dominated by rare propagation events, as in non-linear and active media, or more generally in other fields of research.
\end{bf}

\begin{figure*}%
	\centering
	\includegraphics{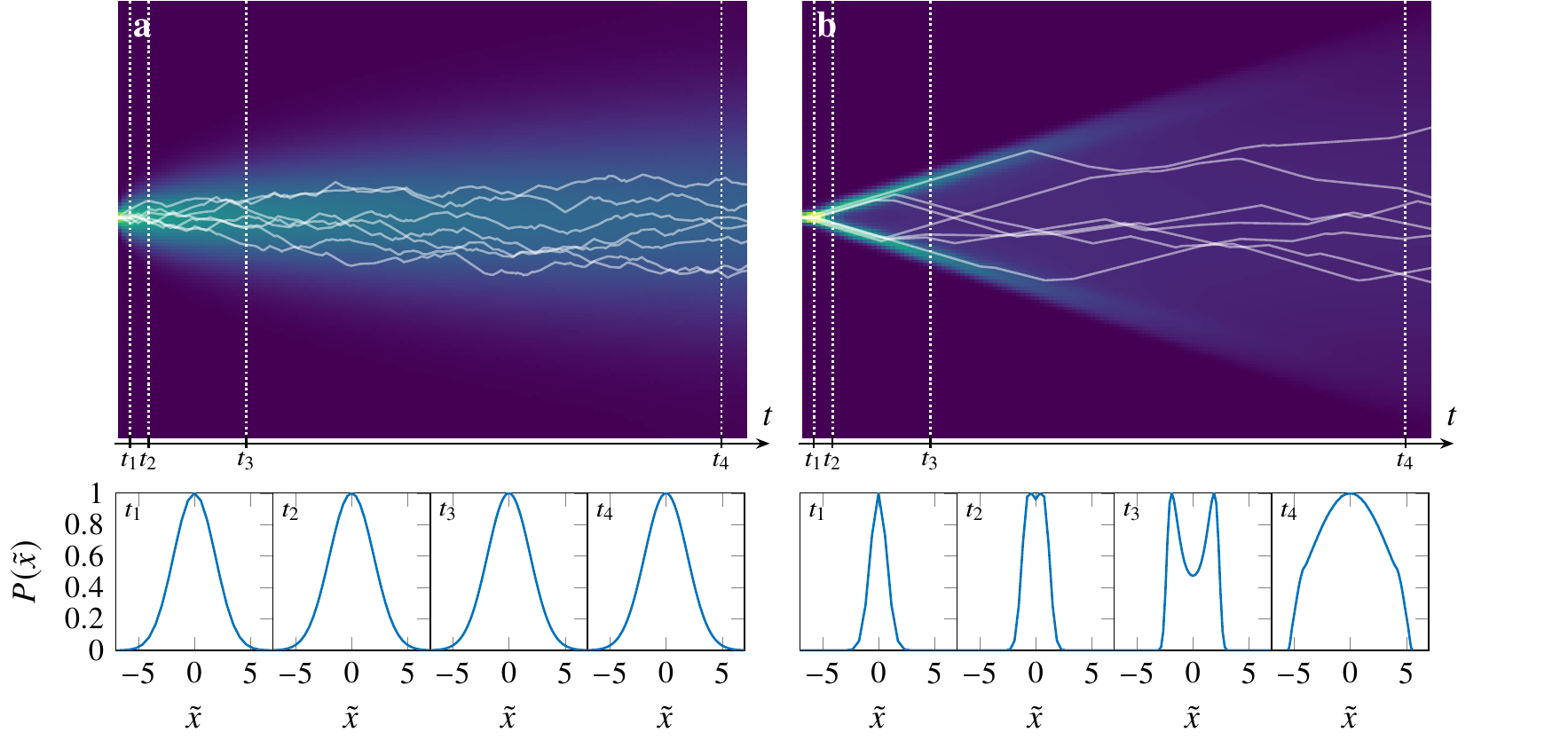}
	\caption{\textbf{Rescaling properties of strongly and weakly self-similar transport processes.} Probability density distributions for a numerical 1D propagation example. A few space-time trajectories are superimposed for illustrative purposes. Cross-cuts of the normalized distributions at different delays are shown in the lower panels. a) In the case of strongly self-similar processes, the normalized spatial profiles can be rescaled onto a single profile at all times, due to the fact that all spatial moments follow the same $q/\nu$ scaling law. b) For more general weakly self-similar processes, no simple rescaling operation can be applied, as the spatial profiles exhibit different functional shapes at different times.}
	\label{fig:SSSvsWSS}
\end{figure*}

The classification of anomalous transport regimes is historically based on the power-law scaling behavior of the mean squared displacement: we call ``normal'' a transport process whose variance grows linearly with time, and ``anomalous'' everything else \cite{metzler2014anomalous}. This classification scheme has obvious limitations, and is unable to tell apart the onset of different propagation regimes\cite{munoz2021objective}.
For this purpose, a richer description can be envisioned based on the scaling behavior of all moments $\expval{\abs{x}^q} \propto t^{\gamma_q}$, rather than just the variance. Studying the power-law growth of different moments of displacement enables a more correct and general classification of all possible transport regimes.
This can be done by building the full ``moment scaling spectrum'' of the process, which is obtained by considering the power-law exponents $\gamma_q$ associated to different $q$-th moments, and evaluating the \emph{self-similarity} of their collective scaling \cite{castiglione1999strong, andersen2000simple, ferrari2001strongly}.
In particular, the transport process is termed \emph{strongly self-similar} if all moments $q$ follow the same scaling law $\gamma_q = q/\nu$ for a given constant $\nu$ (hence $\gamma_q$ grows linearly). Two simple cases exhibiting strong self-similarity are ballistic propagation and ``normal'' diffusion, in which $\nu = 1$ and $\nu = 2$, respectively. Conversely, if $\gamma_q$ is a more complex function of $q$, then the moment scaling spectrum associated the to transport process is said to be \emph{weakly self-similar}.
A qualitative depiction of the typical hallmarks of strongly vs.\ weakly self-similar transport is shown in Fig.\ \ref{fig:SSSvsWSS} for an illustrative 1D case. In the strongly self-similar case, the probability density functions $P(x,t)$ can be collapsed to a single functional shape $F(\tilde{x})$ using a rescaling operation mapping $xt^{-1/\nu} \to \tilde{x}$. In the weakly self-similar case this is not possible, as can be seen intuitively by the fact that the spatial profiles at different times have altogether different shapes.
Based on this concept, the emergence of weak self-similarity has been explored for a variety of similar stochastic models \cite{andersen2000simple, artuso2003anomalous, armstead2003anomalous, vollmer2021displacement}, but experimental studies in this field are still scarce\cite{gal2010experimental, krapf2016strange} due to the lack of a flexible platform where the properties of these transport phenomena can be consistently observed and tuned. Additionally, it should be noted that possible examples of weak self-similarity reported to date arise simply from the presence of residual ballistic tails, meaning that the only form of weak self-similarity reported so far is that resulting by the combination of two separate self-similar processes, rather than a full breakdown distributed across the moment scaling spectrum.

In this work, we show that it is possible to investigate these general transport regimes using light in disordered photonic materials as experimental platform. In particular, we performed transmission studies on optical transport through thin disordered films. By performing a multiple moment scaling analysis of the spatio-temporal transmitted intensity profiles, we show that the hallmarks of non self-similar transport can arise even in surprisingly simple systems. Light transport experiments are traditionally a fertile playground for the study of novel transport regimes, often allowing to cast non-trivial parallels between the propagation dynamics in seemingly unrelated research fields. In most cases, however, peculiar structural and/or optical properties must be purposely introduced in the scattering materials in order to observe these phenomena, such as tailored spatial correlations\cite{bertolotti2010engineering, shi2013amorphous, froufe2017band, dalnegro2017fractional, sellers2017local, jeon2017intrinsic, vynck2021light, aubry2020experimental, patsyk2020observation, yu2021engineered} or wavefront-shaped illumination conditions\cite{rotter2017light, bohm2018situ, yilmaz2019transverse}. For instance, transient anomalous transport regimes were previously observed in structures with fractal heterogeneous inclusions \cite{barthelemy2010role, buonsante2011transport, savo2014walk, svensson2014light}, in agreement with the common belief that only trivial transport regimes are associated to simpler scattering media.
Due to how we typically characterize anomalous transport phenomena, however, several systems exhibiting interesting propagation dynamics may have been overlooked so far.

In the context of light -- and wave transport in general -- perhaps the simplest configuration that can be considered is that of a plane-parallel slab with homogeneous and isotropic disorder. The optical properties of this prototypical model system are determined by the ratio between its physical thickness and its transport mean free path, also known as the optical thickness $L/l_\text{t}$.
As the optical thickness decreases, the validity of the common diffusive approximation breaks down due to the relative increase of a ballistic contribution\cite{lemieux1998diffusing, zhang1999wave, boguna1999persistent, svensson2013exploiting, elaloufi2004diffusive}. Nonetheless, even in this case, a small fraction of light will still spend a long time in the turbid film, undergoing a multiple scattering transport process which may exhibit a multi-scaling behavior. Previous numerical evidence was presented suggesting that light transport in this configuration may be characterized by exceedingly long steps between consecutive scattering events \cite{pattelli2016diffusive}, which may be associated to the onset of previously overlooked transport dynamics.

To investigate this transport regime experimentally, we realize samples with different scattering strengths and use a transient-imaging apparatus capable of recording the spatial and temporal evolution of the intensity profiles based on an optical gating scheme \cite{pattelli2016spatio}. In the experiment, the time-resolved transmittance of a \SI{150}{\femto\second} probe pulse through the scattering sample is sampled via non-linear sum-frequency generation with a collimated gate pulse impinging on a non-linear crystal. The probe and gate beam wavelengths are \SI{1525}{\nano\meter} and \SI{820}{\nano\meter}, respectively. A CCD camera collects the transmitted profiles at different delays, from which the spatial moments can be directly calculated.
The experimental samples consist of free-standing scattering films made of a dilute dispersion of TiO\textsubscript{2} nanoparticles in a UV-cured transparent polymer matrix (see Supplementary Information). Three samples with the same thickness ($L=\SI{100 +- 1}{\micro\meter}$) but different particle density have been fabricated, up to a maximum volume concentration of about \SI{3}{\percent}.

\begin{figure*}
	\centering
	\includegraphics{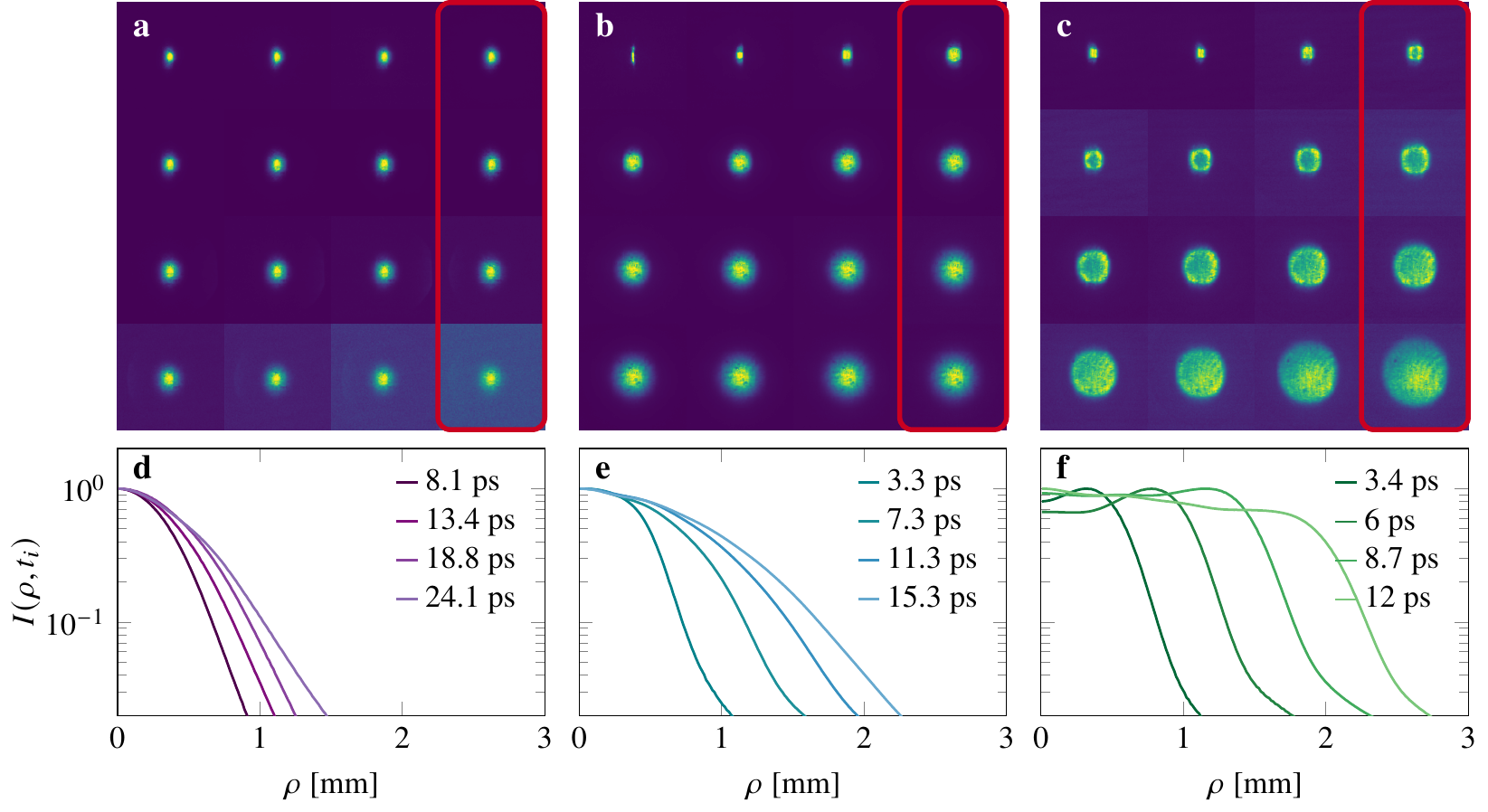}
	\caption{\textbf{Spatio-temporal intensity distributions.} a-c) Transmitted profiles recorded at different times $t_i$ for three \SI{100}{\micro\meter}-thick samples with different turbidity (samples A, B and C in order of decreasing turbidity from left to right). Each frame covers the same area of \num{7.4} $\times$ \SI{7.4}{\milli\meter\squared} and is normalized to its maximum value. Time delays between frames are $\Delta t = \SIlist{1.33;1.0;0.67}{\pico\second}$, respectively, for the three cases. d-f) Representative radially averaged profiles corresponding to the highlighted frames. A time-evolving ring-shaped intensity profile is visible for the least scattering sample C.}
	\label{fig:montages_prof}
\end{figure*}

For each sample -- referred to as sample A, B and C in order of decreasing turbidity -- multiple measurements of the transmitted intensity were recorded and averaged over different positions to obtain the incoherent transmitted intensity profiles at different times $t_i$ (Fig.\ \ref{fig:montages_prof}a-c).
Due to the isotropic structure of the scattering samples, a radial average around the illumination axis can be performed to obtain the profiles $I(\rho, t_i)$ (Fig.\ \ref{fig:montages_prof}d-f). Notably, the profile shape of the instantaneous spatial intensity distributions is independent of the presence of spurious absorption or intensity drifts of the laser sources, which would only be able to modulate their amplitude without affecting the spectrum of their spatial moments. For the same reason, the moments of transmittance profiles recorded using different integration times can also be directly compared without further manipulation.

\begin{figure}
	\centering
	\includegraphics{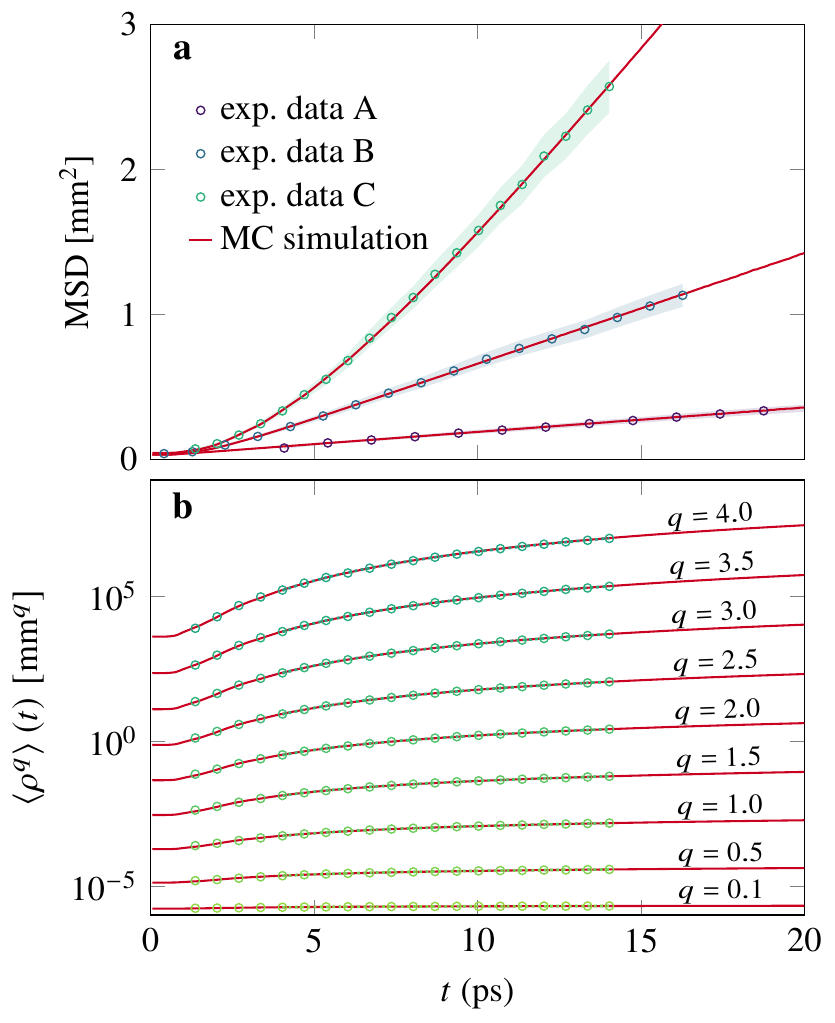}
	\caption{\textbf{Monte Carlo characterization of the scattering samples.} (a) Temporal evolution of the experimental mean squared displacement. Fitting each curve with a corresponding MC simulation returns their associated transport mean free path values. Shaded areas represent $1 \sigma$ confidence intervals.
	(b) Temporal evolution of a subset of the moments of displacement ($0.1 \leq q \leq 4$) for sample C, fitted with eq.\ \eqref{eq:PLfit} for $t > l_\text{t}/v$ (dotted) to retrieve $\gamma_q$. The output of the corresponding MC simulations is also shown (solid, red).}
	\label{fig:MSW}
\end{figure}

The asymptotic growth rate of the second moment (variance) of the transmittance profiles is typically considered as a direct measurement of the diffusion rate in a scattering system\cite{pattelli2016spatio, cobus2022crossover}. We use this observable to retrieve the effective transport mean free path of the considered samples by fitting a Monte Carlo (MC) simulation to the experimental data for each sample (Fig.\ \ref{fig:MSW}a). The analysis returned $l_\text{t}$ values of \SI{53 \pm 4}{\micro\meter}, \SI{195 \pm 11}{\micro\meter} and \SI{630 \pm 30}{\micro\meter} for samples A, B and C, respectively, corresponding to optical thicknesses of \num{1.9 \pm 0.3}, \num{0.51 \pm 0.05} and \num{0.158 \pm 0.017}.

The experimental measurements reveal that the observed asymptotic transverse propagation rate for these samples is much larger than that expected from the $l_\text{t}$ values retrieved by MC fitting. In other words, even after several scattering events, the transverse intensity profiles expand more quickly than what could be expected from the nominal transport mean free path.
The asymptotic values for the mean squared displacement growth rate associated to samples A, B and C are \SI{1.687e4}{\meter\squared\per\second}, \SI{7.624e4}{\meter\squared\per\second} and \SI{2.714e5}{\meter\squared\per\second}, respectively, which in all three cases exceed the $4D$ value associated to the slab geometry (with $D = v l_\text{t}/3$ as the nominal diffusion constant).
For the least scattering sample C, the measured expansion rate is even larger than the $6D$ value expected for the mean squared displacement growth in an unbounded three-dimensional medium\cite{liemert2022analytical}, showing that the observed in-plane diffusion enhancement can even exceed this limit in optically thin geometries.

A more quantitative analysis can be performed by considering the spectrum of all $q$\textsuperscript{th} moments at different times, defined as
\begin{equation} \label{eq:qmomentcont}
	\expval{\rho^q}(t) = \frac{\int_{0}^{\infty}\rho^q I(\rho,t)\rho \dd{\rho}}{\int_{0}^{\infty}I(\rho,t)\rho \dd{\rho}}.
\end{equation}
for a generic intensity distribution $I(\rho,t)$, with $q$ in the positive real numbers.
The growth rate of each moment can be fitted using a power-law model
\begin{equation} \label{eq:PLfit}
	\expval{\rho^q}(t) \sim (t-t_q)^{\gamma_q},
\end{equation}
where $\gamma_q$ is the exponent associated to the $q$\textsuperscript{th} moment and $t_q$ is a temporal offset which can be included to account for the early ballistic transient.

\begin{figure}
	\centering
	\includegraphics{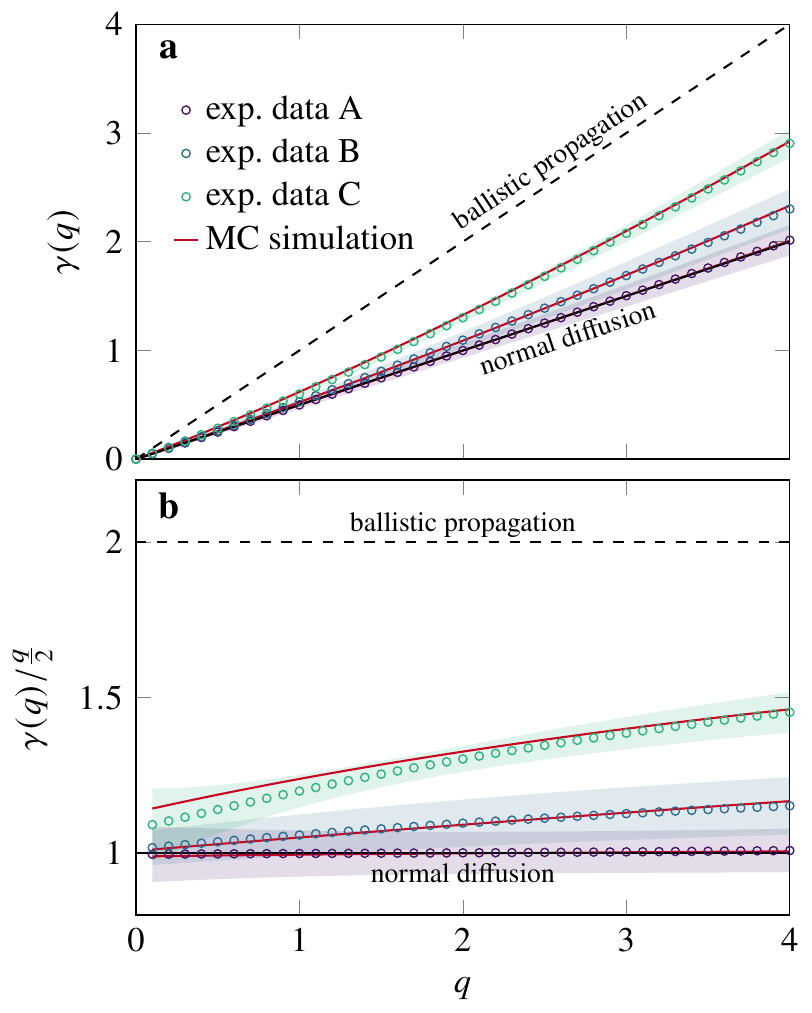}
	\caption{\textbf{Moment scaling spectra analysis.} (a) Experimental and simulated moment scaling spectra, (b) same curves normalized to strong self-similar normal diffusion. The solid and dashed black lines represent the case of strong self-similar normal diffusion and ballistic transport, respectively. Shaded areas in both graphs represent $1 \sigma$ confidence intervals.}
	\label{fig:MSS}
\end{figure}

The moment growth rates were fitted with eq.\ \eqref{eq:PLfit} in a time range excluding the early quasi-ballistic transient ($t > l_\text{t}/v$). The resulting moment scaling spectra are shown in Fig.\ \ref{fig:MSS}a, together with the corresponding results obtained for MC simulations performed with the previously determined $l_\text{t}$ values.
The varying degree of self-similarity exhibited by the three different samples is highlighted in Fig.\ \ref{fig:MSS}b, where the moment scaling spectrum is plotted normalized to the strongly self-similar case ($\gamma(q) = q/2$).
At an optical thickness of \num{1.9}, sample A shows a moment scaling spectrum which is consistent with a normal and strongly self-similar transport regime. On the other hand, samples B and C exhibit hallmarks of both anomalous (super-diffusive) and weakly self-similar transport. In contrast with previous examples, the observed breakdown of self-similarity is spread over a broad range of moments and is not originating from ballistic tails, nor from the presence of engineered heterogeneities.
Indeed, the anomalous transport regime reported here is of a fundamentally different type compared to previously reported examples, such as those associated with L\'{e}vy glasses\cite{barthelemy2008levy, savo2014walk, bertolotti2010engineering}, or quenched scatterer distributions\cite{burioni2010levy, barthelemy2010role, svensson2014light}. In fact, the samples that we have studied are homogeneously disordered, and the effect that we observe persists even when averaging over different disorder configurations, as independently confirmed by annealed-disorder MC simulations. 
These results show that different normal and anomalous transport regimes can be obtained using a very simple material platform and fabrication procedure, comprising both strongly self-similar normal and anomalous diffusion, as well as weakly self-similar propagation.

Due to the finite size of real samples and the dynamic onset of the multiple scattering regime in the slab geometry, different transport regimes are expected to be found at different time scales. We rely on large statistics ($10^{12}$ trajectories) numerical calculation to study how the observed transport regimes evolve asymptotically for $t \to \infty$ for the least scattering configuration (sample C).
By doing so, we find that the transmitted intensity profiles evolve from a super-diffusive and weakly self-similar behavior to a moment spectrum compatible with strongly self-similar normal diffusion (see Supplementary Information). For a sample with a nominal optical thickness of \num{0.158}, this transition to the strongly self-similar regime eventually occurs at a delay $\SI{> 60}{\pico\second}$, corresponding to a total path length of $\SI{> 1}{\centi\meter}$, well within the multiple scattering regime ($\approx \num{20}l_\text{t}$). 
On the other hand, the in-plane diffusion rate enhancement with respect to the nominal transport mean free path persists also in the limit for $t \to \infty$, i.e., even after the transport regime has converged to its asymptotic and strongly self-similar final form.

In conclusion, we studied the propagation of light in optically thin media with homogeneous disorder, performing a full moment scaling spectrum analysis which revealed the presence of anomalous transport transient regimes which can be super-diffusive and non self-similar.
Despite its unique ability to identify different transport regimes and reveal their properties, the application of this kind of analysis for the experimental study of transport phenomena is still sporadic \cite{munoz2021objective, wang2012brownian}. Up to date, a weakly self-similar mobility of polymer particles inside living cancer cells has been associated with the presence of active transport mechanisms\cite{gal2010experimental, goldstein2013origin}, and to the occurrence of bulk-mediated long jumps for particles being adsorbed and desorbed on planar lipid bilayers\cite{krapf2016strange}. Notably, all these cases only showed weak self-similarity of the same bi-linear type, where the low moments describe the expansion rate of the central region of the spatial profile, while higher moments exhibit a ballistic scaling for the tails.

In thin scattering membranes, light intensity propagates realizing a more complex (and general) breakdown of self-similarity, which changes continuously along the moment spectrum, at different delays and for different degrees of turbidity in the scattering slab geometry, in excellent agreement with numerical calculations. Therefore, these physical systems can serve as a versatile platform for the study of a broad range of general properties of transport phenomena, such as hallmarks of non-Gaussian diffusion\cite{wang2012brownian}, extreme events\cite{vezzani2020rare}, transient return probabilities\cite{levernier2021universality} or visitation statistics\cite{regnier2023universal}, to name a few.

The effects that we have described are observed after several transport mean free paths, meaning that they are not associated to ballistic propagation nor to a particular illumination condition, and could be used for instance to check whether the invariance property for the average pathlength in scattering media is compatible with, or even requires, the onset of the observed transverse transport enhancement to remain valid up to arbitrary precision\cite{martelli2021verification}. Similarly, this class of non self-similar transport arises without the need to postulate the presence of specific conditions such as heterogeneous disorder, active transport mechanisms, or jumps mediated through different embedding media -- which suggests that the occurrence of weakly or altogether non self-similar transport may be more frequent than previously assumed.

Considering the rich picture that we obtained by performing a quantitative analysis of the moment scaling spectrum of light transport in such a simple and general system, it can be further envisioned that even more complex dynamics will be revealed when applying this analysis to materials with a more complex structure, possibly allowing us to reveal the emergence of distinct transport regimes that would not be correctly identified with the traditional variance-based classification scheme.

\section{End notes}

\subsection{Data availability}
The data are available upon request.

\subsection{Acknowledgments}
L.P. acknowledges financial support by the European Union's NextGenerationEU Programme with the I-PHOQS Research Infrastructure [IR0000016, ID D2B8D520, CUP B53C22001750006] ``Integrated infrastructure initiative in Photonic and Quantum Sciences'', and NVIDIA Corporation for the donation of the Titan X Pascal GPU used for this research. L.P. wishes to thank F. Martelli and S. Lepri for discussion. E.P. thanks S. Donato for his help with sample preparation. D.S.W. acknowledges financial support from the European Union's Horizon 2020 research and innovation programme under FET-OPEN Grant Agreement No.\ 828946 (PATHOS).

\subsection{Author Contributions}
E.P. fabricated the samples, performed the experiment, ran the simulations and analyzed the data. G.M. and L.P. developed the simulation software. L.P. conceived the project, directed the research, and wrote the first draft of the manuscript. All authors discussed the results and commented on the manuscript.

\subsection{Additional information}
Supplementary Information is available for this paper.

\subsection{Corresponding author}
Correspondence and requests for materials should be addressed to Lorenzo Pattelli.

%

\onecolumngrid
\clearpage
\begin{center}
	\textbf{\large Supplementary Information: Breakdown of self-similarity in light transport}
\end{center}

\setcounter{equation}{0}
\setcounter{figure}{0}
\setcounter{table}{0}
\setcounter{page}{1}
\makeatletter
\renewcommand{\theequation}{S\arabic{equation}}
\renewcommand{\thefigure}{S\arabic{figure}}
\renewcommand{\thetable}{S\arabic{table}}
\renewcommand{\bibnumfmt}[1]{[S#1]}
\renewcommand{\citenumfont}[1]{S#1}

\section{Supplementary Methods: Samples fabrication} \label{sec:samplefabrication}
	
Three samples were fabricated with a thickness of $L=\SI{100 +- 1}{\micro\meter}$ and different scattering densities, targeting optical thickness values between \num{0.1} and \num{10}, i.e., spanning across the ballistic-to-diffusive transition.
	
The samples are composed of a mixture of a transparent UV-curable glue (\textsc{Norland Optical adesive} 65) and Titanium dioxide nanoparticles (\textsc{Tioxide RX-L}) with an average diameter of \SI{280}{\nano\meter}. We start with preparing various mixtures by varying the mass fraction between TiO\textsubscript{2} and the glue, as determined using a precision balance. For dilute samples, according to the independent scattering approximation, increasing the concentration of nanoparticles has the effect of increasing the optical thickness. The fabrication steps are illustrated in Fig.\ \ref{fig:samplefabrication}:
\begin{figure}[h]
	\centering
	\includegraphics[width=\textwidth]{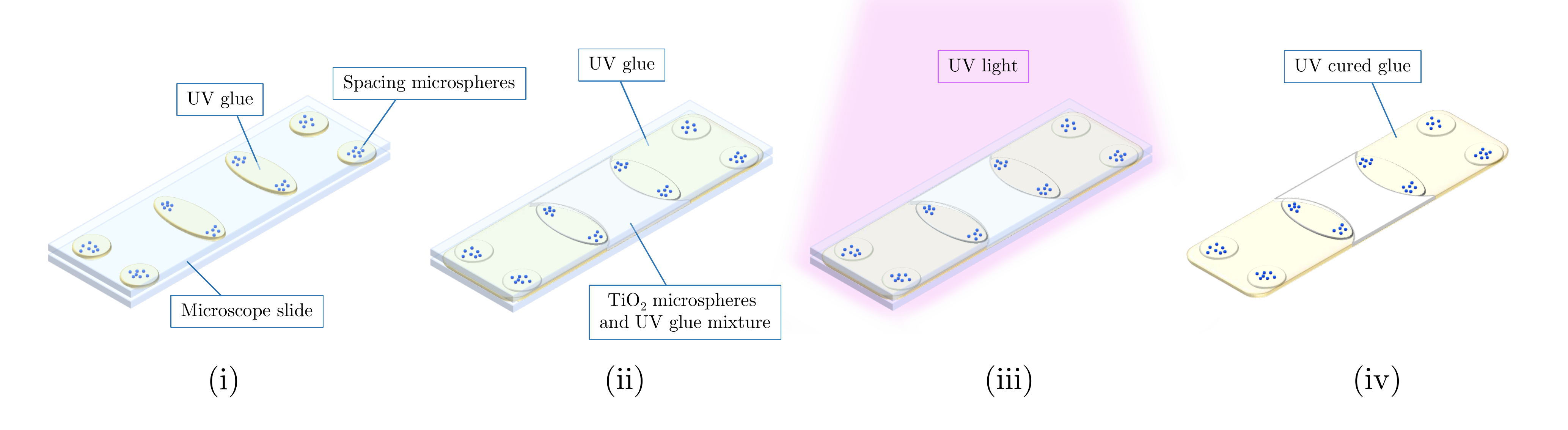}
	\caption{Sample fabrication steps including: (i) fixed-thickness glass cell fabrication, (ii) scattering paste infiltration, (iii) UV curing and (iv) detachment of the sample from the glass slides.}
	\label{fig:samplefabrication}
\end{figure}
	
\begin{enumerate}[label=(\roman*)]
	\item Two microscope slides are blade-coated with a thin layer of water-soluble Polyvinyl alcohol (PVA), to help the subsequent removal of the sample from the glass cell after UV curing. A few \textsc{NIST}-calibrated glass microspheres with a diameter of \SI{100}{\micro\meter} are placed on the edge of the lower glass slide to act as spacers. A second glass slide is placed in contact with the microspheres forming a controlled gap between the two slides, which are glued together via UV curing.
	\item A scattering paste with the desired concentration of TiO\textsubscript{2} and glue is mixed using a magnetic stirrer for at least \num{2} hours to obtain a homogeneous mixture. This mixture is then infiltrated between the two microscope slides by capillarity.
	\item The sample is exposed to a \SI{35}{\watt} UV lamp for \num{6} hours to cure the scattering mixture (the nominal curing time of the transparent glue is around 5 minutes, but the presence of TiO\textsubscript{2} nanoparticles slows down the curing process).
	\item The sample is placed in water to dissolve the PVA layers and allow the release of the sample as a free-standing film. The sample is then cured again under UV light for a few hours on each side to ensure its complete polymerization.
\end{enumerate}

This process results in flexible, free-standing slabs with a central scattering section of at least \SI{5}{\centi\meter\squared} where the transport measurement are performed. Beside the scattering area, two transparent regions are used to measure the exact sample thickness optically through time-of-flight measurement of the round-trip time of internal reflections, and to precisely determine the absolute origin of the time-domain axis.

The properties of the three fabricated samples are reported in Tab.\ \ref{tab:samples}. The TiO\textsubscript{2}/UV glue volume ratio is estimated from the mass ratio assuming spherical nanoparticles and a TiO\textsubscript{2} specific gravity of \SI{3.55}{\gram\per\cubic\micro\meter}.

\setlength{\tabcolsep}{1em}
\begin{table}[h]
\centering
\caption{}	
\begin{tabular}{c c S S c}
	\toprule
	\addlinespace 
		Sample	& Spacer size &	{TiO\textsubscript{2}/UV-cured} & {TiO\textsubscript{2}/UV-cured} & Optical \\
		& (\SI{}{\micro\meter}) & {glue mass ratio} & {glue volume ratio} &  Thickness \\
	\midrule
	A  & \num{100} & 0.1 & 0.034 & $1.9 \pm 0.3$\\
	B  & \num{100} & 0.02 & 0.0068 & $0.51 \pm 0.05$\\ 
	C  & \num{100} & 0.005 & 0.0017 & $0.158 \pm 0.017$\\
	\bottomrule
\end{tabular}
\label{tab:samples}
\end{table}

\section{Supplementary Discussion: Asymptotic analysis} \label{sec:asymptoticanalysis}

\begin{figure}[h]
	\centering
	\includegraphics{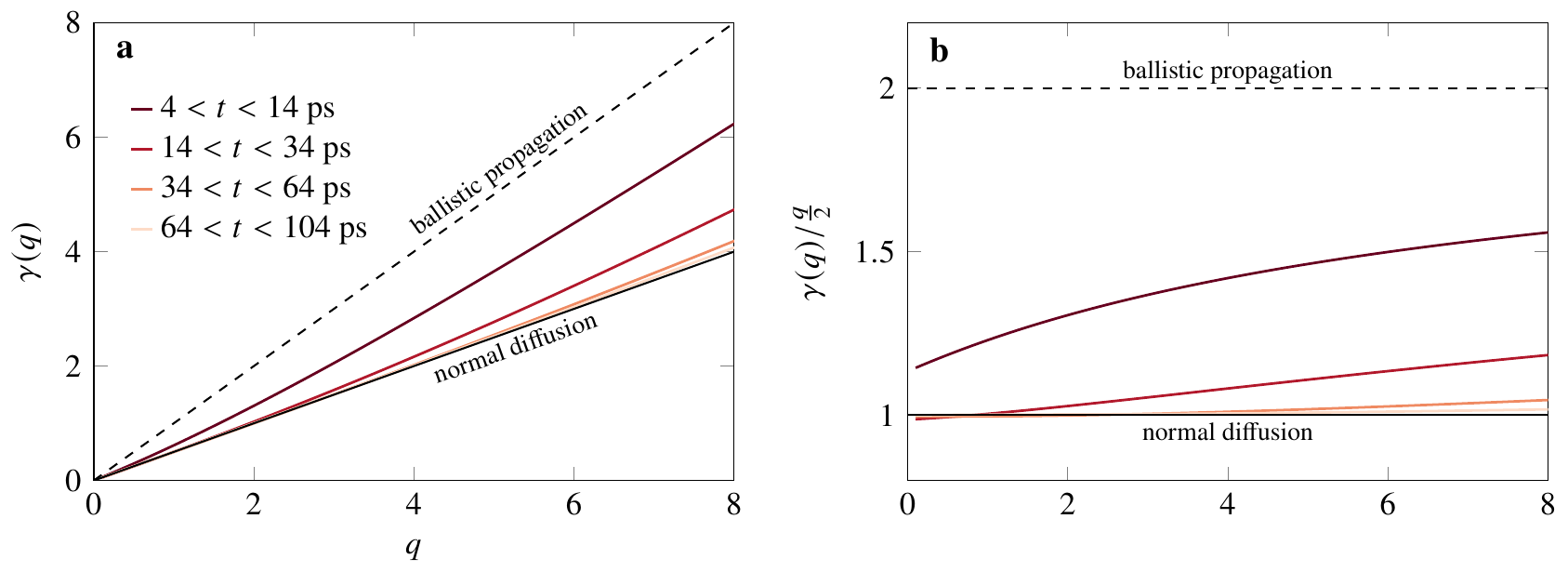}
	\caption{\textbf{Asymptotic evolution of moment scaling spectra.} (a) Moment scaling spectra for sample C calculated for non-overlapping fitting time intervals at increasing delays, (b) same curves normalized to strong self-similar normal diffusion. The solid and dashed black lines represent the case of strong self-similar transport, respectively for normal diffusion and ballistic transport.}
	\label{fig:MSSasymptotic}
\end{figure}

Here we report the results of the moment analysis performed at different time windows, relative to the MC simulation for a sample with nominal optical thickness of \num{0.158} (Sample C). The growth rates of the moments of displacement $0.1 < q < 8$ were fitted with eq.\ \eqref{eq:PLfit} for increasingly delayed time ranges, resulting in the moment scaling spectra shown in Fig.\ \ref{fig:MSSasymptotic}. A progressive transition from a super-diffusive and weakly self-similar behavior to strongly self-similar normal diffusion is observed even at this low optical thickness after a delay of approximately \SI{60}{\pico\second}.

\end{document}